\documentstyle[preprint,aps,epsfig]{revtex}
\widetext
\draft
\tighten

\begin{document}

\title{Unusual formations of the free
electromagnetic field in vacuum}

\bigskip

\author{{\bf Andrew E. Chubykalo} and {\bf Augusto Espinoza}
}

\address {Escuela de F\'{\i}sica, Universidad Aut\'onoma de Zacatecas \\
Apartado Postal C-580\, Zacatecas 98068, ZAC., M\'exico\\
e-mail: {\rm  achubykalo@yahoo.com.mx}}

\date{\today}

\maketitle

%  \date{December 12, 1995}

\baselineskip 7mm

\begin{abstract}
It is shown that there are exact solutions of the free Maxwell equations
(FME) in vacuum  allowing an existence of stable spherical
formations of the free magnetic field  and ring-like formations of
the free electric field.  It is detected that a form of these spheres and
rings does not change with time in vacuum. It is shown that these
{\it convergent} solutions are the result of an interference of some {\it
divergent} solutions of FME. One can surmise that these electromagnetic
formations correspond to Kapitsa's hypothesis about interference origin
and a structure of {\it fireball}.
\end{abstract}


%\newpage

\section{Introduction}

The generally accepted opinion exists that solutions of the free Maxwell
equations (FME) are well studied and do not boil down to any surprises.
Nevertheless, we will show in the next Sections, such
mathematically well-known solutions (see, e.g., [1], where general
solutions of the Maxwell equations was obtained) lead, however, to the
existence of rather unusual and unexpected electromagnetic formations in
vacuum such as closed spherical magnetic surfaces ({\it without} electric
field on this surface and where the magnetic field is tangential and its
intensity depends on time) and ring-like formations of the electric field
({\it without} magnetic field in all points of the ring and where the
electric field is tangential and depends on time).  We will also show that
these formations do not change their form with time in vacuum.

\section{An unusual solution  of the free Maxwell equations}

We found that a certain class of exact solutions
of FME
\begin{eqnarray} && {\rm div}\,{\bf
E}=0,\\ && {\rm rot}\,{\bf E}=-\frac{1}{c}\frac{\partial{\bf B}}{\partial
t},\\ && {\rm div}\,{\bf B}=0,\\ && {\rm rot}\,{\bf
B}=\frac{1}{c}\frac{\partial{\bf E}}{\partial t}.
\end{eqnarray}
exists which has some unexpected characteristics. The present work is
devoted  to the research of the such solutions.

We shall look for solutions of the system FME as follows:
\begin{equation}
{\bf E}({\bf r},t)={\bf e}({\bf r})\psi(t)\qquad {\rm
and}\qquad {\bf B}({\bf r},t)={\bf b}({\bf r})\chi(t).
\end{equation}
where  $\psi(t)$ and $\chi(t)$ are some
functions of time, vectors ${\bf e}$ is a polar vector and ${\bf b}$ is an
axial one.

And so substituting (5) in FME we obtain:

\begin{eqnarray}
&& {\rm div}\,{\bf e}=0,\\
&& {\rm rot}\,{\bf e}=-\frac{1}{c}\frac{\chi^{\prime}}{\psi}
{\bf b}\\
&& {\rm div}\,{\bf b}=0,\\
&& {\rm rot}\,{\bf
b}=\frac{1}{c}\frac{\psi^{\prime}}{\chi}{\bf e}.
\end{eqnarray}
It is obvious that these equations are consistent if and only if
\begin{equation}
-\frac{\chi^\prime}{\psi}=\Omega_1
\qquad{\rm and}\qquad
\frac{\psi^\prime}{\chi}=\Omega_2,
\end{equation}
where index \{$ ^{\prime}$\} means a derivative with respect to
time, $\Omega_1$ and $\Omega_2$ are arbitrary constants.

In order to obtain solutions of this system with three constants only, and
to obtain sinusoidal solutions, we propose that $\Omega_1=\Omega_2=\Omega$.
Thus, the general solution of the system (10) is

\begin{equation} \chi(t)=
{\cal A}\cos(\Omega t-\eta)\qquad{\rm and}\qquad \psi(t)=
{\cal A}\sin(\Omega t-\eta),
\end{equation}
where ${\cal A}$ and $\eta$ are arbitrary constants, and equations for
${\bf e}$ and ${\bf b}$ become:

\begin{equation} \nabla\times{\bf
e}=\frac{\Omega}{c}{\bf b}\qquad{\rm and}\qquad \nabla\times{\bf
b}=\frac{\Omega}{c}{\bf e}.
\end{equation}

In order to solve this system, let us at first note that formally summing
two equations (12) we obtain
\begin{equation}
\nabla\times({\bf e}+{\bf
b})=\frac{\Omega}{c}({\bf e}+{\bf b})\qquad{\rm or} \qquad\nabla\times{\bf
a}=\frac{\Omega}{c}{\bf a}.
\end{equation}
So, at first we resolve Eq.(13) with respect to ${\bf a}$, and then we
obtain from the vector ${\bf a}$ (which, obviously, has no  polarity)
the polar vector ${\bf e}$ and the axial vector ${\bf b}$. Actually, one
can express polar and axial parts of any vector without polarity, in
general, as follows:

\begin{equation}
{\bf e}({\bf
r})=\frac{1}{2}\Bigl[{\bf a}({\bf r})-{\bf a}(-{\bf r})\Bigr]
\end{equation}
and

\begin{equation} {\bf b}({\bf r})=\frac{1}{2}\Bigl[{\bf
a}({\bf r})+{\bf a}(-{\bf r})\Bigr]
\end{equation}
Now, if we calculate a rotor of both parts of equations (14), (15) one can
be satisfied that the system (12) is fulfilled:

\begin{equation}
\nabla\times{\bf e}({\bf r})=\frac{1}{2}\Bigl[\nabla\times{\bf a}({\bf r})
-\nabla\times{\bf a}(-{\bf r})\Bigr] =
\frac{1}{2}\Bigl[\frac{\Omega}{c}{\bf a}({\bf r})+\frac{\Omega}{c}{\bf
a}(-{\bf r})\Bigr]=\frac{\Omega}{c}{\bf b}({\bf r})
\end{equation}
and

\begin{equation}
\nabla\times{\bf b}({\bf r})=\frac{1}{2}\Bigl[\nabla\times{\bf a}({\bf r})
+\nabla\times{\bf a}(-{\bf r})\Bigr] =
\frac{1}{2}\Bigl[\frac{\Omega}{c}{\bf a}({\bf r})-\frac{\Omega}{c}{\bf
a}(-{\bf r})\Bigr]=\frac{\Omega}{c}{\bf e}({\bf r}).
\end{equation}
Here we take into account that after inverting the coordinates, the
equation $\nabla\times{\bf a}({\bf r})=\frac{\Omega}{c}{\bf a}({\bf r})$
becomes $-\nabla\times{\bf a}(-{\bf r})=\frac{\Omega}{c}{\bf a}(-{\bf
r})$. Thus, one can see that if we find ${\bf a}$ as a solution of Eq.(13)
it means that we find ${\bf e}$ and ${\bf b}$ as solution of the system
(12).

The equation (13) was already solved in the literature (see, e.g.  [2,3]).

And so, the solution of  Eq.(13)  in the spherical
system of coordinates is\footnote{${\cal D}$ is a dimension
constant $[{\cal D}]={\tt M}^{1/2}{\tt L}^{5/2}{\tt T}^{-1}$}:

\begin{equation}
{\bf a}={\cal D}\left\{\frac{2\alpha}{r^3}\cos\theta\right\}{\bf e}_r+
{\cal D}\left\{\frac{\gamma}{r^3}\sin\theta\right\}{\bf e}_{\theta}+
{\cal D}\left\{\frac{\Omega\alpha}{cr^2}\sin\theta\right\}{\bf
e}_{\varphi}.
\end{equation}

Finally, separating vectors ${\bf e}$ and ${\bf b}$ we obtain the solution
of the system (12) expressed by components (cartesian and spherical ones):

\begin{equation} {\bf e}={\cal D}\left\{-\frac{\alpha \Omega y}{c
r^3},\quad \frac{\alpha \Omega x}{c r^3},\quad 0\right\}=
\frac{\Omega\alpha\sin\theta}{cr^2}{\cal D}{\bf e}_\varphi
\end{equation}
and

\begin{equation}
{\bf b}={\cal D}\left\{\frac{\beta xz}{r^5},\quad \frac{\beta
yz}{r^5},\quad \frac{2\alpha}{r^3}-\frac{\beta(x^2+y^2)}{r^5}\right\}=
\frac{2\alpha\cos\theta}{r^3}{\cal D}{\bf e}_r+
\frac{\gamma\sin\theta}{r^3}{\cal D}{\bf e}_\theta,
\end{equation}
where
$$
\alpha=\cos\left(\frac{\Omega r}{c}-\delta\right)+\frac{\Omega r}{c}\sin
\left(\frac{\Omega r}{c}-\delta\right).
$$

$$
\beta=3\alpha-\frac{\Omega^2r^2}{c^2}\cos\left(\frac{\Omega
r}{c}-\delta\right)\qquad {\rm and}\qquad\gamma=\beta-
2\alpha.
$$

Let us now write  the solution (5) in the explicit form, taking into
account Eqs.(11), (19) and (20):

\begin{equation}
{\bf E}=\left[\frac{\Omega\alpha\sin\theta}{cr^2}{\cal D}{\bf
e}_\varphi\right] \sin(\Omega t-\eta) \end{equation} and \begin{equation}
{\bf B}= \left[\frac{2\alpha\cos\theta}{r^3}{\cal D}{\bf e}_r+
\frac{\gamma\sin\theta}{r^3}{\cal D}{\bf e}_\theta\right]
\cos(\Omega t-\eta),
\end{equation}
where $\delta$ and  $\eta$  are  arbitrary constants.

It follows from the solutions (21), (22) that the necessary (not
sufficient!) condition in order for these solutions to not diverge in $r=0$
is:
$$
\alpha(0)=\left\{\cos\left(\frac{\Omega r}{c}-\delta\right)+
\frac{\Omega r}{c}\sin\left.\left(\frac{\Omega r}{c}-\delta\right)\right\}
\right|_{r=0} = 0 \qquad\Longrightarrow
$$
$$
\cos\delta=0\qquad\Longrightarrow\qquad
\delta=(n+\frac{1}{2})\pi,\;{\rm where}\; n=0,\pm 1,\pm 2, \ldots
$$

In order to satisfy oneself that the solutions (21), (22) converge, one can
calculate the following limits\footnote{we calculate these limits
expanding $\alpha$ and $\gamma$ in series of powers of $r$.} for $\delta =
\frac{\pi}{2}$:

\begin{equation}
\lim\limits_{r\rightarrow 0}\frac{\alpha}{r^2}=0;\qquad
\lim\limits_{r\rightarrow
0}\frac{\alpha}{r^3}=\frac{\Omega^3}{3c^3};\qquad
\lim\limits_{r\rightarrow 0}\frac{\gamma}{r^3}=-\frac{2\Omega^3}{3c^3},
\end{equation}
and corresponding limits for ${\bf E}$, ${\bf B}$ and the energy density
$w=\frac{E^2+B^2}{8\pi}$ are:

\begin{equation}
\lim\limits_{r\rightarrow 0}{\bf E}=0;\qquad
\lim\limits_{r\rightarrow 0}{\bf B}=
\frac{2{\cal D}\Omega^3\cos(\Omega t)}{3c^3}{\bf k};\qquad
\lim\limits_{r\rightarrow 0}w=\frac{{\cal D}^2\Omega^6}{18\pi c^6}
\cos^2(\Omega t),
\end{equation}
where ${\bf k}$ is {\it Z-ort} of the cartesian system.

The constant $\eta$ just defines an initial wave phase  of the fields
${\bf E}$ and ${\bf B}$. So without loss of generality we can just write
one non-divergent solutions  for  $\delta=\frac{\pi}{2}$, $\eta=0$:

\begin{equation}
{\bf E}=
{\cal D}\left[
\frac{\alpha\Omega\sin\theta}{cr^2}
{\bf e}_\varphi
\right]
\sin(\Omega t);\quad
{\bf B}=
{\cal D}\left[
\frac{2\alpha\cos\theta}{r^3}
{\bf e}_r+
\frac{\gamma\sin\theta}{r^3}
{\bf e}_\theta
\right]
\cos(\Omega t),
\end{equation}
where
\begin{equation}
\alpha=-\frac{\Omega r}{c}\cos\left(\frac{\Omega r}{c}
\right)+
\sin\left(\frac{\Omega r}{c}\right)\qquad{\rm and}\qquad \gamma=
\alpha-\frac{\Omega^2r^2}{c^2}\sin \left(\frac{\Omega r}{c}\right).
\end{equation}
Note that the solution (25) can be found directly from the general
solution of the Maxwell equations obtained by G. Mie [1]

Generally speaking, taking into account  (19) and (20) one can see that
an infinity divergent solutions for ${\bf E}$ and ${\bf B}$ exists as well
as convergent ones. Curiously enough,  but the convergency of these
solutions is defined by the constant $\delta$. Solutions converge in $r=0$
if and only if

\begin{equation}
\delta=(n+\frac{1}{2})\pi,\;{\rm where}\; n=0,\pm 1,\pm 2,\; \ldots
\end{equation}

We show that the convergent solution (25)\footnote{We designate it  by
${\bf E}_c$ and ${\bf B}_c$ in this section.} is represented as an
interference of divergent ones.

After simple algebraic transformation one can represent the solution (25)
as a superposition of two waves spreading in opposite directions in every
point:

\begin{equation}
{\bf E}_c={\bf E}_{(\rightarrow)} +{\bf E}_{(\leftarrow)}\qquad{\rm
and}\qquad {\bf B}_c={\bf B}_{(\rightarrow)} +{\bf B}_{(\leftarrow)},
\end{equation}
where
\begin{equation}
{\bf E}_{(\rightarrow)} = \frac{\Omega\sin\theta}{2cr^2}
\left[\cos\left(\frac{\Omega r}{c}-\Omega t\right)+
\frac{\Omega r}{c}\sin\left(\frac{\Omega r}{c}-\Omega t\right)
\right]{\cal D}{\bf e}_\varphi,
\end{equation}

\begin{equation}
{\bf E}_{(\leftarrow)} = -\frac{\Omega\sin\theta}{2cr^2}
\left[\cos\left(\frac{\Omega r}{c}+\Omega t\right)+
\frac{\Omega r}{c}\sin\left(\frac{\Omega r}{c}+\Omega t\right)
\right]{\cal D}{\bf e}_\varphi,
\end{equation}

\begin{eqnarray}
\lefteqn{
{\bf B}_{(\rightarrow)} = \frac{\cos\theta}{r^3}
\left[\sin\left(\frac{\Omega r}{c}-\Omega t\right)-
\frac{\Omega r}{c}\cos\left(\frac{\Omega r}{c}-\Omega t\right)
\right]{\cal D}{\bf e}_r +} \nonumber\\
& & +\frac{\sin\theta}{2r^3}\left[
-\frac{\Omega r}{c}\cos\left(\frac{\Omega r}{c}-\Omega t\right)+
\left(1-\frac{\Omega^2 r^2}{c^2}\right)
\sin\left(\frac{\Omega r}{c}-\Omega t\right)\right]{\cal D}{\bf e}_\theta,
\end{eqnarray}

\begin{eqnarray}
\lefteqn{
{\bf B}_{(\rightarrow)} = \frac{\cos\theta}{r^3}
\left[\sin\left(\frac{\Omega r}{c}+\Omega t\right)-
\frac{\Omega r}{c}\cos\left(\frac{\Omega r}{c}+\Omega t\right)
\right]{\cal D}{\bf e}_r +} \nonumber\\
& & +\frac{\sin\theta}{2r^3}\left[
-\frac{\Omega r}{c}\cos\left(\frac{\Omega r}{c}+\Omega t\right)+
\left(1-\frac{\Omega^2 r^2}{c^2}\right)
\sin\left(\frac{\Omega r}{c}+\Omega t\right)\right]{\cal D}{\bf e}_\theta,
\end{eqnarray}

Labourless calculation shows also that

\begin{equation}
{\bf E}_{(\rightarrow})=\frac{1}{2}(-{\bf E}_d+{\bf E}_c);
\qquad {\bf E}_{(\leftarrow})=\frac{1}{2}({\bf E}_d+{\bf E}_c)
\end{equation}
and

\begin{equation}
{\bf B}_{(\rightarrow})=\frac{1}{2}(-{\bf B}_d+{\bf B}_c);
\qquad {\bf B}_{(\leftarrow})=\frac{1}{2}({\bf B}_d+{\bf B}_c),
\end{equation}
where ${\bf E}_d, {\bf B}_d$ are divergent solutions of the system
(1)-(4):

\begin{equation}
{\bf E}_d=
{\cal D}\left[
\frac{\alpha_d\Omega\sin\theta}{cr^2}
{\bf e}_\varphi
\right]
\cos(\Omega t);\quad
{\bf B}_d=
{\cal D}\left[
\frac{2\alpha_d\cos\theta}{r^3}
{\bf e}_r+
\frac{\gamma_d\sin\theta}{r^3}
{\bf e}_\theta
\right]
\sin(\Omega t),
\end{equation}
here
$$
\alpha_d=\cos\left(\frac{\Omega r}{c}
\right)+
\frac{\Omega r}{c}\sin\left(\frac{\Omega r}{c}\right)\qquad{\rm
and}\qquad \gamma_d= \alpha-\frac{\Omega^2r^2}{c^2}\sin \left(\frac{\Omega
r}{c}\right).
$$

It is obvious that the functions ${\bf E}_{(\rightarrow)},{\bf
E}_{(\leftarrow)},{\bf B}_{(\rightarrow)},{\bf B}_{(\leftarrow)}$ diverge
in $r=0$ and they are also solutions of FME.

\section{Stable electromagnetic spheres and rings in vacuum as a
consequence of the solution (25)}

As we will show below the solution (25) of FME leads to an
existence of unusual spherical formations of the free
electromagnetic field.

\subsection{Some details of the energy distribution in the field (25)}

Let us write, after some transformations, the expression for the energy
density for the solution (25). One can show that the energy density
contains both the time-independent part and time-dependent one:

\begin{eqnarray}
w=\frac{E^2+B^2}{8\pi} & = & \frac{{\cal D}^2}{16\pi}
\left\{
\frac{\Omega^2\alpha^2}{c^2r^4}\sin^2\theta+
\left[
\frac{4\alpha^2}{r^6}\cos^2\theta+\frac{\gamma^2}{r^6}\sin^2\theta
\right]
\right\}+  \nonumber\\
& & +\frac{{\cal D}^2}{16\pi}
\left\{
\left[
\frac{4\alpha^2}{r^6}\cos^2\theta+\frac{\gamma^2}{r^6}\sin^2\theta
\right]
-\frac{\Omega^2\alpha^2}{c^2r^4}\sin^2\theta
\right\}
\cos(2\Omega t).
\end{eqnarray}

Let us find from (36) the {\it locus} where $w$ does not
depend on $t$.  It is obvious that the {\it loci} are

1) along the axis $Z$ in the points where
$\tan\left(\frac{\Omega z}{c}\right)=\frac{\Omega z}{c}$
($\theta=0,\pi;\alpha=0$);

2) at surfaces where $r$ satisfies the equation
$\gamma^2=\alpha^2\left(\frac{\Omega^2 r^2}{c^2}-4\cot^2\theta\right)$.
One can see the cross-section of these surfaces in Fig.3
(discontinuous curves)\footnote{All figures in this work were performed in
the program ``Mathematica-4.0".}

Now we calculate the electromagnetic energy ${\cal E}$ within a sphere of
the radius $R$ with the centre in the coordinate origin:

\begin{equation}
{\cal E}_\oplus = \int\limits_{0}^{R}\!\!\!dr\!
\int\limits_{0}^{\pi}\!\!d\theta\!\!\int\limits_{0}^{2\pi}
\!\!\!d\varphi\;r^2\sin\theta\;w(r,\theta,\varphi,t)
= {\cal E}(R)+ {\cal
E}(R,t), \end{equation}
where

\begin{equation}
{\cal E}(R)= \frac{{\cal D}^2}{6R^3}\left[\frac{\Omega^4R^4}{c^4}
-\frac{\Omega^2R^2}{c^2}\sin^2\left(\frac{\Omega R}{c}\right)-
\alpha^2\right],
\end{equation}

\begin{equation}
{\cal E}(R,t)=-\frac{{\cal D}^2}{6R^3}\alpha\gamma
\cos(2\Omega t).
\end{equation}
Here $\alpha=-\frac{\Omega R}{c}\cos\left(\frac{\Omega R}{c}\right)+
\sin\left(\frac{\Omega R}{c}\right)$ and
$\gamma=\alpha-\frac{\Omega^2R^2}{c^2}\cos \left(\frac{\Omega R}{c}\right)$

One can show from Eq.(39) that electromagnetic energy within spheres of the
radiuses $R$ which are solutions  of the equations\footnote{it
follows from $\alpha=0$ and $\gamma=0$ correspondingly.}

\begin{equation}
\tan\left(\frac{\Omega R}{c}\right)=\frac{\Omega R}{c}
\end{equation}
or
\begin{equation}
\tan\left(\frac{\Omega R}{c}\right) =\frac{\frac{\Omega
R}{c}}{1-\frac{\Omega^2R^2}{c^2}}, \end{equation} does not change with
time.

Note that every root of Eq. (40) is placed at the number line between two neighbour roots of Eq. (41) and vice versa. One can  show that a distance between
theseneighboring spherical surfaces tends to $\frac{c\pi}{2\Omega}$ when
$R\rightarrow\infty$.  Let us also direct attention to an interesting fact
that at the surfaces of the spheres of the radius (40)  only the
magnetic field is present, and the electric field at these surfaces does
not exist. It follows directly from Eq.(25) for $\alpha=0$.

\subsection{Analysis of the Poynting vector's field corresponding to the
wave field (25)}

The Poynting's vector corresponding to the wave field (25) is

\begin{equation}
{\bf S}=\frac{c}{4\pi}{\bf E}\times{\bf B}=\frac{{\cal D}^2}{8\pi}
\left[
\frac{\Omega\alpha^2\sin(2\theta)}{r^5}{\bf e}_\theta-
\frac{\Omega\alpha\gamma\sin^2\theta}{r^5}{\bf e}_r
\right]
\sin(2\Omega t).
\end{equation}

Let us calculate the total momentum and the angular momentum of the
electromagnetic field (25) within a sphere of the arbitrary radius $r$
with the center in the coordinate origin. Because the Poynting's vector is
proportional to the vector of the density of momentum in the same point we
can just calculate the integral of the Poynting's vector over volume of
the sphere.

It is easy to calculate this integral if we express spherical system {\it
orts} by cartesian system {\it orts}:
$$
{\bf e}_r={\bf i}\sin\theta\cos\varphi+{\bf j}\sin\theta\sin\varphi+{\bf
k}\cos\theta\quad{\rm and}\quad
{\bf e}_\theta={\bf i}\cos\theta\cos\varphi+{\bf
j}\cos\theta\sin\varphi-{\bf k}\sin\theta.
$$
Thus, integrating (42) over volume of the sphere we obtain:

\begin{equation}
\int\!\!\!\int\!\!\!\int {\bf S}\; r^2\sin\theta\,dr\,d\theta\,d\varphi=
-\frac{{\cal D}^2 4\pi \Omega\sin(2\Omega t)}{32\pi}{\bf
k}\int\left.\frac{\alpha^2}{r^3}\sin^4\theta\right|_0^{\pi}\;dr=0.
\end{equation}

It means that the total momentum of the electromagnetic field (25) in
volume bounded by an arbitrary  sphere with a center in the coordinate
origin {\it is zero} at any time. Analogically one can show that the total
angular momentum of this field configuration is zero.

Let us now find the {\it loci} where Poynting's vector is zero at any
instant of time.  It follows from Eq.(42) that conditions when Poynting's
vector is zero are:
\begin{equation}
\alpha^2\sin(2\theta)=0
\qquad{\rm and}\qquad
\alpha\gamma\sin^2\theta=0.
\end{equation}
From the first equation of the conditions (44) we have the following
possibilities:

(i) $\alpha=0$. This automatically satisfies both conditions (44). From
$\alpha=0$ we obtain the equation
\begin{equation}
\tan\left(\frac{\Omega r}{c}\right)=\frac{\Omega r}{c}.
\end{equation}
Hence, the {\it loci} for the case (i) are spheres whose radiuses satisfy
Eq.(45).

(ii) $\sin(2\theta)=0$. It means that $\theta$ can be 0,
$\frac{\pi}{2}$ or $\pi$.

(ii-1) If $\theta$ is 0 or $\pi$, in this case both equations fulfill the
conditions (44). So the {\it locus} is axis $Z$.

(ii-2) If $\theta=\frac{\pi}{2}$, this gives us two possibilities in order
to satisfy the conditions (44): {\it either} $\alpha=0$ (it is the case
(i), see above) {\it or} $\gamma=0$. From the last we have

\begin{equation}
\tan\left(\frac{\Omega r}{c}\right)=
\frac{\frac{\Omega r}{c}}{1-\frac{\Omega^2r^2}{c^2}}.
\end{equation}
So the {\it loci} corresponding the case $\theta=\frac{\pi}{2}$, and
$\gamma=0$ are rings at the plane $XY$ with radiuses satisfying Eq.(46).
Note that in all points of these {\it rings} the magnetic field is {\it
zero}.

Now  we consider spheres whose equators are mentioned {\it
rings}. These spheres are defined by the condition $\gamma =0$. One can
see from Eq.(42) that at these surfaces the Poynting's vector in all
points has  tangential components only. Due to this fact the
conservation of the energy within spheres of the radiuses (41) becomes
more clear.

Thus, adjusted total in looking-for the {\it loci} where the Poynting's
vector for the field (25) is zero in any instant of time is:

{\it Locus} 1: Axis $Z$. We call this axis  {\it magnetic axis} because
an electric field does not exist there.

{\it Locus} 2: Rings at the plane $z=0$ with radiuses satisfying
Eq.(46). We call these rings  {\it electric rings} because a magnetic
field does not exist there.

{\it Locus} 3: Spheres with centers in the origin with radiuses satisfying
Eq.(45). We call these spheres  {\it magnetic spheres} because the
electric field does not exist on them.

In order to elucidate better the results of the last analysis, let us
adduce the graphic (FIG.1) where the distribution of the Poynting's vector
field is shown.

\begin{figure}
\caption{Poynting's vector field distribution for given instant of
time in the plane $x=0$, $Y$-axis is the abscissa and $Z$-axis is
the ordinate.  Here $c=1, \omega=1$.}
\end{figure}

We consider this distribution, for example, in the plane
$x=0$ (because of the axial symmetry of the energy density and
energy-flux density distribution it is sufficient to  consider this
cross-section only).

We call spheres whose equator is the {\it electric ring} E-sphere.
We call the {\it magnetic spheres} M-spheres. In the Fig.1
one can see the vertical {\it magnetic axis} (coinciding with the
$Z$-axis), the first E-sphere, the first M-sphere  and the second
E-sphere  in the given instant of time. Within E-spheres  the total
electromagnetic energy conserves because the energy-flux vector at the
surface of this sphere has tangential component only.  The energy
transfers   along this surface from pole to equator ({\it electric ring})
and after a certain period\footnote{This period is defined by the
function $\sin(2\Omega t)$ from Eq.(42).} of time does reverse movement.
Within the first E-sphere the energy transfers from the {\it magnetic
axis} to the {\it electric ring} and after a certain time returns.

The Poynting vector is zero in every point of the first M-sphere  so the
energy within this sphere conserves too. One can see that the energy
transfers from the surface of the first {\it magnetic sphere} to the {\it
electric rings} of the first and the second E-spheres. Analogical
exchange of the energy takes place between next E- and M-spheres.

We once more emphasize that the  Poynting's vector field takes
opposite direction  with time, due to the existence of the function
$\sin(2\Omega t)$ in Eq.(42).

For more demonstrativeness we adduce here the graphic (FIG.2) of
cross-section of the Poynting's vector field in the plane $z=0$.

\begin{figure}
\caption{Poynting's vector field distribution for given instant of
time in the plane $z=0$, $X$-axis is the abscissa and $Y$-axis is
the ordinate.}
\end{figure}

At last we adduce here the common graphic (FIG.3) of cross-section
($x=0$) of the surfaces where the energy density is constant and first
M-sphere, first and second E-spheres.

\begin{figure}
\caption{Cross-section of the surfaces of the constant energy density
and first M-sphere and first E-spheres in the plane $x=0$, $Y$-axis is
the abscissa and $Z$-axis is the ordinate Here $c=1, \omega=1$.}
\end{figure}

We emphasize that  these surfaces do not deform, do not displace and do
not rotate with time in vacuum.

\section{Discussion}

Thus, we obtained a stationary {\it free} electromagnetic field which can
be consequence of some interference processes. Why one can speak here
about interference? Actually, we see that in this electromagnetic
formation,  surfaces  (discontinuous curves in Fig.3)  and points (in
$Z$-axis) where the energy density is {\it constant} exist. From this one
can surmise that these surfaces and points are nodes of wave. It is
well-known also that standing electromagnetic waves are a result of
interference processes.

Of course, the solution of the {\it free} Maxwell equations corresponding
to these ball-like electromagnetic formations was obtained for vacuum. But
if we call to mind that in air, the values $\varepsilon=1,\; \mu=1$
we can be practically sure that the solution (25) is valid for  air,
taking into account that air does not have free charges and currents. So it
is easy to draw an analogy between our solution and Kapitsa's hypothesis
about the {\it interference nature} of ball lightning [4]. Actually,
the electric field of electromagnetic waves which ``voyage" within
M-spheres and especially the electric field of the aforementioned {\it
electric rings} have to ionize the air converting it to plasma.  A size of
the critical region of ionization is defined by the radius of the {\it
magnetic sphere}, in which a density energy is still adequate for ionizing
air.  This ultimate {\it magnetic sphere} in turn plays the role of a
magnetic trap for plasma confinement. One can indeed see from Eq.(36) that
the energy density within the magnetic spheres decreases as
$\frac{1}{r^2}$.  It means that at a certain distance the energy density
is less than the critical value which is necessary to ionize the air. This
condition has to define the radius of the ultimate magnetic sphere within
which conditions of the ionization still exist. Taking into account this
limited value of the radius of this ultimate magnetic sphere  one can
speak about the fireballs.

It goes without saying that it is just our hypothesis, but the analogy
between Kapitsa's idea and our {\it ball-like} solutions doubtless takes
place. It should also be stated that other ball-like stable formations in
the radiation field  were obtained in the paper ``Is there yet an
explanation of ball lightning?" by G.H. Arnhoff [5] and in the paper
``Ball lightning as a force-free magnetic knot" by A. F. Ra\~nada {\it et
al} [6] (see also [7]).  It follows from these works that the
electromagnetic energy contained in a spherical volume, cannot escape (the
energy corresponding to our solutions behaves in the same way). According
to [5] outside of this volume there is  only an quasi-electrostatic field,
rotating with constant angular velocity about the axis (in this point our
and Arnhoff's solutions are different). In turn A.F. Ra\~nada {\it et al}
[6,7] proposed ball-like electromagnetic formations as  a solution based
on the idea of the ``electromagnetic knot", an electromagnetic field in
which any pair of magnetic lines or any pair of electric lines form a link
- a pair of linked curves.

Thus the famous hypothesis of the Nobel prizewinner P.L.Kapitsa that
fireballs (or balls lightning) are standing electromagnetic waves of
unusual configuration as a result of some {\it interference} process from
the day of its formulation (in 1955) never (to the present day) got a
theoretical (mathematical) support. One can see that our work first gives
a theoretical support to this hypothesis.

In a subsequent work we are going to research the process of the genesis
of these unusual electromagnetic formations.

And in conclusion we just note that A.O. Barut was right when claimed
that  ``{\it Electrodynamics and the classical theory of fields
remain very much alive and continue to be the source of inspiration for
much of the modern research work in new physical theories}" [8].

\acknowledgments

The authors would like to express their gratitude to Prof. of School of
Physics  of Zacatecas University V. Dvoeglazov and especially to Profs. V. Onoochin and
G. Kotel'nikov from Russia for their discussions and critical comments.
We would also like to thank Annamaria D'Amore for revising the manuscript.

\end{document}